\documentstyle[12pt,cite]{article}                                        
\catcode`\@=11                                                       
\begin{document}                                                     

\renewcommand{\theequation}{\thesection.\arabic{equation}}        
\newcommand{\mysection}[1]{\setcounter{equation}{0}\section{#1}}  

\def\ee{\end{equation}}
\def\be{\begin{equation}}
\def\la{\label}
\def\dxi{\partial_\xi}
\def\D{{\cal D}}
\def\sin{{\rm sin}}
\def\cos{{\rm cos}}
\def\f{{\bf \Phi}}
\def\v{\varphi}
\def\O{{\bf {\Omega}}_\nu}

\begin{titlepage} 

\hfill{SISSA / 114 / 98/ EP }

\vspace{2 cm }

\centerline{\Huge  Holography and CFT on Generic Manifolds}

\vspace{4 cm }
\centerline{\large Giulio Bonelli
\footnote{e-mail:{\tt bonelli@sissa.it}}}

\centerline{\it SISSA/ISAS and INFN Trieste}

\centerline{\it Via Beirut 2-4; Trieste, Italy}
   
\vspace{3 cm}
\centerline{\large ABSTRACT} 
\vspace{.5 cm}

In this paper it is shown how the AdS/CFT correspondence extends to 
a more general situation in which the first theory is defined
on $(d+1)$-dimensional manifold $\tilde M$ defined as the filling
in of a compact $d$-dimensional manifold $M$.
The stability of the spectral correspondence 
mass/conformal-weight under such geometry changes is also proven.
  
\end{titlepage}

\newpage

\section{Introduction}

During last year there has been a big effort in understanding AdS/CFT
correspondence \cite{maldacena,gubser,witten,ferrara}.
Among others, the question arose of how to obtain the CFT
on the boundary of the AdS space for various field theories
(see also \cite{henningson,volovich,dhoker,mathur}). In these papers
a method was proposed to demonstrate also a spectral correspondence between
the mass spectrum of the theory on the AdS space and the conformal 
weight spectrum of the CFT on the boundary $\partial$AdS.
This correspondence is of course crucial in establishing the full
String/SYM mapping and is at the basis of this realization of the 
t'Hooft-Susskind holography \cite{thooft,sussho}.

A natural interesting question concerns possible generalizations of this 
mapping for string vacua of the form AdS$\times$Q with Q not a sphere,
in order to obtain on the other side CFTs with less supersymmetry
\cite{klebwit,G/H,kehagias}.

In this letter we address the problem of proving an analogous 
statement for geometries which are not AdS, but more general.
We will consider the geometry of a space $\tilde M$, with dimension
$d+1$, obtained by filling up a generic compact manifold $M$, with 
dimension $d$; we will define it with an appropriate metric such that the 
conformal transformations on $M$ are included as isometries on 
$\tilde M$.
In this construction $M$ plays the role of a boundary compactified 
space-time.

The general correspondence between the two theories can be expressed as
\be
\int_{\tilde\phi\vert_{\partial\tilde M=M}=\phi_0}
\D\left[\tilde\phi\right]
e^{iS_{\tilde M}\left(\tilde g,\tilde\phi\right)}=
\int\D\left[\phi\right]
e^{iS_M \left(g,\phi\right)
+i\int_M\sqrt{g}{\cal O}_\delta[\phi]\cdot\phi_0}
\la{prima}\ee
(in a regularized sense), where $S_M$ and $S_{\tilde M}$ are the actions 
of the two related theories on $M$ and $\tilde M$ , $g$ and $\tilde g$ are 
the respective metrics, $\phi$ and $\tilde\phi$ represent 
fields on the manifolds and ${\cal O}_\delta[\phi]$ a generic composite 
field of some conformal dimension $\delta$.
Notice that, in general, for interacting bulk
theories the problem arizes of obtaining sharp conditions on it
so that on $\partial\tilde M=M$ one is left possibly with a local
field theory.

The above program is viable by generalizing the AdS approach 
\cite{gubser,witten}. In the next section we will study the case of the
massive interacting scalar boson and the free massive spinor
to explain the general method.
In the rest of the paper the correspondence QFT on $\tilde M$ versus
CFT on $M=\partial\tilde M$ is explained together with the proof of 
stability of the spectral map.

Given the above situation, one could obtain potentially interesting string 
or M-theory
vacua of the form $\tilde M\times Q$ and speculate on an extension of
the string/SYM correspondence. The study of the properties of
this class of vacua is not performed in this letter.

\section{Some explicit calculations}

Let $(M,g)$ be a compact riemannian manifold of dimension $d$ and let 
$(\tilde M,\tilde g)$ the riemannian manifold 
$\tilde M={\bf R}_+\times M$ equipped with the metric
$$ d \tilde s ^2 = {d\xi^2 + ds^2 \over \xi^2}$$ where
$ds^2 = g_{ab} dx^adx^b$ is the metric on $M$ and $\xi$ is a coordinate
on ${\bf R_+}$.

Notice that each isometry of $(M,g)$ is also an isometry of
$(\tilde M,\tilde g)$. More interesting is the fact that each
dilatation of $(M,g)$, $ds^2\,\to\, r^2 ds^2$, is also an isometry
of $(\tilde M,\tilde g)$, once $\xi\,\to\, r\xi$.

In the following we will exploit a technique to discharge to
the harmonic analysis on the manifold $M$ all its peculiar
geometrical data. This is obtained via transferring all the 
amplitudes calculations in a Fourier transformed form.

\subsection{The free massive scalar boson}

In this section we analize the free massive scalar boson theory.

Let $\Delta$ and $\tilde\Delta$ be the scalar Laplacians on $M$ and 
$\tilde M$ respectively. By definition, the following relation
holds 
\be
\tilde\Delta=\xi^2 \Delta -\xi^{d+1}\dxi\xi^{1-d}\dxi
\, .
\la{dtd}\ee

Let now $m$ be the mass of the scalar $\tilde\phi$: its equation 
of motion is
\be
\left(\tilde\Delta +m^2\right)\tilde\phi=0\, . 
\la{eom}\ee
We write its solution as an integral 
in terms of the would-be boundary value $\phi_0$ 
$$
\tilde\phi(\xi,x)=\int_M d^dy \sqrt{g(y)} \phi_0(y) \Gamma(\xi|x,y)\, .
$$
The Green function is to be expanded as
$$
\Gamma(\xi|x,y)=\sum_{\lambda} \gamma(\xi|\lambda)\psi_\lambda(x)
\psi_\lambda(y)\, ,
$$
where we denoted by $\{\psi_\lambda\}$ a complete orthonormal set of
solutions of $\Delta \psi_\lambda =\lambda \psi_\lambda$ on $M$.

Using (\ref{dtd}) we obtain from (\ref{eom}) the following equation for 
$\gamma(\xi|\lambda)$ $$
\left(\xi^2 \lambda -\xi^{d+1}\dxi\xi^{1-d}\dxi
\right)\gamma + m^2 \gamma=0
$$
which can be shown to be reducible to the modified Bessel equation
of order $\nu=\sqrt{(d/2)^2+m^2}$.
The solution which is regular at $\xi\sim\infty$ and nicely matches
at $\lambda\sim 0$ is then
$$
\gamma(\xi|\lambda)= c' \left({\sqrt\lambda \over 2}\right)^\nu
\xi^{{d\over 2}} 
K_\nu\left(\xi\sqrt{\lambda}\right)\, ,
$$
where $K_\nu$ is the relative modified Bessel function of order 
$\nu$ and $c'$ is a normalization.

We can now perform the calculation of the regularized classical action for 
the field configuration $\tilde\phi$ on $\tilde M_\epsilon=[\epsilon,\infty)
\times M$ 
$$
\tilde I_\epsilon \left(\tilde\phi\right)=
{1\over2}\int_{\tilde M_\epsilon}
d^{d+1}\tilde x\sqrt{\tilde g}
\left(
\tilde g ^{AB}\partial_A\tilde\phi\partial_B\tilde\phi
+m^2\tilde\phi^2
\right)
= 
{1\over2}\int_{\tilde M_\epsilon}
d^{d+1}\tilde x\partial_A\left(
\sqrt{\tilde g}
\tilde g ^{AB}\tilde\phi\partial_B\tilde\phi
\right)\, .
$$
This reduces, using Stoke's theorem, to
$$
{1\over2}\int_M d^dx \left(
\tilde\phi\vert_{\xi=\epsilon}
\epsilon^{1-d}\sqrt{g}
\dxi\tilde\phi\vert_{\xi=\epsilon}
\right)=
{1\over2}\sum_\lambda \hat\phi_0(\lambda)^2\epsilon^{1-d}
\left[\gamma\dxi\gamma\vert_{\xi=\epsilon}\right]\, ,
$$
where $\hat\phi_0(\lambda)=\int_Md^dx\sqrt{g}\psi_\lambda\phi_0$
and where we used completeness and orthonormality of the armonics 
$\{\psi_\lambda\}$.
Extracting the finite part of $\epsilon^{1-d}\gamma\dxi\gamma$ as
$\epsilon\to\, 0$
$$
{\rm fin}\left[\epsilon^{1-d}\gamma\dxi\gamma\right]=
c\left({d\over2}+\nu\right)\lambda^\nu\, ,\quad
c={c'^2\pi\over 2^{2\nu+2}\nu {\rm sin}(\pi\nu)}
$$
we get 
$$
\tilde I_{\rm fin}\left(\tilde\phi\right)=
c\left({d\over2}+\nu\right)
{1\over2}\sum_\lambda\hat\phi_0(\lambda)^2\cdot\lambda^\nu=
$$ $$=
c\left({d\over2}+\nu\right) {1\over2}\int_M
d^dx\, d^dy\sqrt{g}(x)\sqrt{g}(y)
\phi_0(x)\phi_0(y)\Omega_\nu(x,y)
$$
where $\Omega_\nu(x,y)=\sum_\lambda\psi_\lambda(x)\psi_\lambda(y)\cdot
\lambda^\nu$.
This is exactly the generating functional for 2-point functions
of a CFT for a field of conformal dimension $\delta={d\over2}+\nu$. 
In fact, under a scale transform $g_{ab}\,\to\, r^2\, g_{ab}$, we
get $\lambda\,\to\, r^{-2}\lambda$ and 
$\psi_\lambda\,\to\, r^{-d/2}\psi_\lambda$ and then
$\Omega_\nu\,\to\, r^{-d-2\nu}\Omega_\nu$. 

In the case $M={\bf R}^d\cup\{\infty\}$ everything above reduces to the usual
AdS calculation and $\Omega_\nu(x,y)\sim{1\over |x-y|^{d+2\nu}}$.

\subsection{The interacting massive scalar boson}

In this subsection we analyze perturbatively the interacting scalar boson 
which we denote by $\tilde\f$. The perturbative analysis will be performed 
in the classical limit
for the cubic potential ${h\over3!}\tilde\f^3$ in the first order in $h$.
Higher orders in $h$ can be analized in the same way. In the following 
formulas we will omit sometimes the specific indication of higher order 
corrections $o(h)$.

The e.o.m. of the field is 
\be
\left(\tilde\Delta +m^2\right)\tilde\f + {h\over 2}\tilde\f^2=0\, . 
\la{eomi}\ee
We write its solution as an integral 
in terms of the would-be boundary value $\f_0=\phi_0+h\v_0$ 
$$
\tilde\f(\xi,x)=\int_M d^dy \sqrt{g(y)}
\sum_{\lambda} \psi_\lambda(x)
\psi_\lambda(y)\gamma(\xi|\lambda)\left[\phi_0(y)+h g(\xi|\lambda)
\v_0(y)\right]=\tilde\phi_0+h\tilde\v_0
$$
where $g$ is normalized as $g(\xi|\lambda)=1+O(\xi)$ near the boundary.
Eq. (\ref{eomi}) splits in
$$
\left(\tilde\Delta +m^2\right)\tilde\phi_0=0\,;\quad
\left(\tilde\Delta +m^2\right)\tilde\v_0 + {1\over 2}\tilde\phi_0^2=0\, . 
$$
and
we obtain for $g(\xi|\lambda)$ the equation
\be
\partial_\xi\left(\xi^{-d+1}\gamma(\xi|\lambda)^2\partial_\xi g(\xi|\lambda)
\right)\hat\v_0(\lambda)=
-\xi^{-(d+1)}\gamma(\xi|\lambda)\sum_{\lambda'\,\,\lambda''}
\gamma(\xi|\lambda')\gamma(\xi|\lambda'')c_{\lambda\lambda'\lambda''}
\hat\phi_0(\lambda')\hat\phi_0(\lambda'')
\la{solv}\ee
where $c_{\lambda\lambda'\lambda''}\equiv
\int_M d^dx \sqrt{g(x)}
\psi_\lambda(x)\psi_{\lambda'}(x)\psi_{\lambda''}(x)$.

The classical action is evaluated as follows
$$
\tilde I_\epsilon^h \left(\tilde\f\right)=
{1\over2}\int_{\tilde M_\epsilon}
d^{d+1}\tilde x\sqrt{\tilde g}
\left(
\tilde g ^{AB}\partial_A\tilde\f\partial_B\tilde\f
+m^2\tilde\f^2+{h\over3}\tilde\f^3
\right)
= $$ $$ =
{1\over2}
\int_{\tilde M_\epsilon}d^{d+1}\tilde x
\left[
\partial_A\left(
\sqrt{\tilde g}
\tilde g ^{AB}\tilde\f\partial_B\tilde\f
\right)
+\sqrt{\tilde g}\left(
\tilde\f\left(\tilde\Delta +m^2\right)\tilde\f 
+ {h\over 3} \tilde\f^3\right)
\right]=
$$ $$ =
\tilde I_\epsilon \left(\tilde\phi\right)
+h
\int_{\tilde M_\epsilon}d^{d+1}\tilde x
\left[
\partial_A\left({1\over2}
\sqrt{\tilde g}
\tilde g ^{AB}\partial_B(\tilde\phi\tilde\v)
\right)
-{1\over12}
\sqrt{\tilde g}
\tilde\phi^3
\right]
$$ 
and, using the Stokes theorem and (\ref{solv}), we get its finite part
$$
{\rm fin}\left[
\tilde I_\epsilon^h \left(\tilde\f\right)\right]
=
{1\over2}
\int_M d^dx\sqrt{g(x)}
\int_M d^dy\sqrt{g(y)}
\f_0(x)\O(x,y)\f_0(y)
+$$ 
\be
+{h\over3!}
\int_M d^dx\sqrt{g(x)}
\int_M d^dy\sqrt{g(y)}
\int_M d^dz\sqrt{g(z)}
\f_0(x)\f_0(y)\f_0(z)\O(x,y,z)\, ,
\la{yesss}\ee
where 
$$
\O(x,y)= c\left({d\over2}+\nu\right)\Omega_\nu(x,y)
\quad {\rm and}
$$ $$
\O(x,y,z)=b \sum_{\lambda\lambda'\lambda''}
\psi_{\lambda}(x)\psi_{\lambda'}(y)\psi_{\lambda''}(z)
c_{\lambda\lambda'\lambda''}
\left(\lambda\lambda'\lambda''\right)^{\nu/2}
H\left(\lambda,\lambda',\lambda''\right)\, ,
$$
with $b= 5\cdot 2^{-3\nu-1} {c'}^3$ and
$$
H\left(\lambda,\lambda',\lambda''\right)=
\int_0^{+\infty}d\xi \xi^{{d\over2}-1}
K_\nu\left(\sqrt{\lambda}\xi\right)
K_\nu\left(\sqrt{\lambda'}\xi\right)
K_\nu\left(\sqrt{\lambda''}\xi\right)\, .
$$

Using the same rescalings as for the free scalar boson case, 
under the conformal transformation
$g_{ab}\,\to\, r^2 g_{ab}$ we get
$\O(x,y,z)\,\to\, r^{-3\left({d\over2}+\nu\right)}
\O(x,y,z)$ and therefore $\O(x,y,z)$ represents
a good three point function for a conformal field of dimension
$\delta={d\over2}+\nu$.

This shows that under the interaction the mass/conformal weight
correspondence remains stable also in this larger framework
in which the boundary geometry is generic.

Notice that the above results reproduce the interacting scalar boson amplitudes
in the $M={\bf R}^d\cup\{\infty\}$ case. In particular, the coefficient
$H\left(\lambda,\lambda',\lambda''\right)$
reproduces the interaction vertex as calculated with the Witten-diagrams
technique \cite{mathur,dhoker}.

\subsection{The free massive spinor}

In this subsection we analyze the free spinor field which we denote
by $\tilde\psi$. We simplify a little possible questions about the 
harmonic spinor analysis for manifolds with boundary and torsion 
(see for example \cite{peeters} for indication of problems).

The action is
\be
\tilde S_\epsilon\left(\tilde\psi\right)=
\int_{\tilde M_\epsilon}d^{d+1}\tilde x\sqrt{\tilde g(\tilde x)}
\bar{\tilde\psi}(\tilde x)\left(\tilde{\D}-m\right)\tilde\psi(\tilde x)
+\mu\int_{\tilde M_\epsilon}{\cal L}_{\tilde v}\left[
d^{d+1}\tilde x\sqrt{\tilde g(\tilde x)}
\bar{\tilde\psi}(\tilde x)\tilde\psi(\tilde x)
\right]\, ,
\la{spiact}\ee
which is the usual Dirac action augmented by a boundary term which
is the natural covariantization of the analogous term proposed in 
\cite{henningson}
with $\tilde v(\tilde x)=\xi\partial_\xi+v^a(\xi,x)\partial_{x_a}$ a 
vector field of this given form and ${\cal L}_{\tilde v}=i_{\tilde v}\cdot 
\tilde d +\tilde d\cdot i_{\tilde v}$ the Lie derivative.
From this technical point of view, it seems
unnecessary to impose any condition on the components $v^a$ since 
they do not appear in the final result. 
The constant $\mu$ is a weigth for the boundary term.

The Dirac operator $\tilde{\D}$ can be written in the form
$$
\tilde{\D}
=\tilde e^A_M\Gamma_A\left(
\partial^M+{1\over2}\tilde\omega^M_{BC}\Sigma^{BC}\right)=
=\xi\left[
e^a_m\Gamma_a\left(
\partial^m+{1\over2}
\omega^m_{bc}\Sigma^{bc}\right)\right]+
\left(\xi\partial_\xi-{ d \over 2 }\right)
\Gamma_0=
$$
\be
=\xi\D+
\left(\xi\partial_\xi-{ d \over 2 }\right)\Gamma_0
\la{diracop}\ee
where $\tilde e^A_M$ and $\tilde\omega^M_{BC}$ 
[resp. 
$e^a_m$ and $\omega^m_{bc}$] 
are the components
of the inverse vielbein and the spin connection on $\tilde M$
[resp. on $M$].
Let $\{\chi_{q\eta}(x)\}$ be the set of complete orthonormal 
spinors of
the operator $\D$ on $M$, that is
$\D\chi_{q\eta}
=iq\eta\chi_{q-\eta}$ and $\Gamma_0\chi_{q\eta}=\eta\chi_{q\eta}$.

The relevant equation of motion is the Dirac equation
\be
\left(\tilde{\D}-m\right)\tilde\psi(\tilde x)=0
\la{dirac}\ee
and we write its solution as 
\be
\tilde\psi(\tilde x)=\sum_{q,\eta}\beta_\eta(\xi|q)
\chi_{q\eta}(x)
\hat\psi^0_\eta(q)\,,
\la{spide}\ee
where $\hat\psi^0_\eta(q)=\int_Md^dy\sqrt{g(y)}\bar\chi_{q\eta}(y)
\psi^0_\eta(y)$ in terms of the would be boundary spinor
$\psi^0(y)=\psi^0_{+1}(y)+\psi^0_{-1}(y)$.

Substituting (\ref{spide}) in (\ref{dirac}) and imposing square integrability,
we get
\be
\beta_\eta(\xi|q)=c''(q) e^{i\pi\eta/4}
\sin\left(\eta m-{1\over2}\right)
\xi^{{d+1\over2}} 
K_{{1\over2}-\eta m}(q\xi)
\la{spiso}\ee
for $q\not=0$ and $\beta_\eta(\xi|0)\propto\xi^{\eta m+{d\over2}}$.
$\beta_\eta(\xi|q)$ is well defined
at $q=0$ only for the value of $\eta$
dictated by the condition $\eta m\leq 0$ that is, for 
$m\geq 0$, $\eta=-1$. As a consequence it is also determined
$c''(q)=q^{m+{1\over2}}c''$.
We are then forced to kill half the boundary value of the spinor 
from the beginning as 
$$\psi^0_{+1}(x)=0$$

Calculating the classical action from (\ref{spiact}) we get
$$
{\rm fin}\left[
\tilde S_\epsilon\left(\tilde\psi\right)
\right]=
-\mu a \sum_q
\bar{\hat\psi}_{q,-1}
{\hat\psi}_{q,-1} q^{2m}=
$$
\be
=-\int_Md^dx\sqrt{g(x)}\int_Md^dy\sqrt{g(y)}
\bar\psi^0_{-1}(x) \Theta(x,y)\psi^0_{-1}(y) 
\, ,
\la{spiok}\ee
where $a=|c''|^2 {\pi\over2}\cos^2(\pi m)$
and 
$$
\Theta(x,y)=\mu a
\sum_q \chi_{q-1}(x) q^{2m}\bar\chi_{q-1}(y)\,.$$

Notice that (\ref{spiok}) is the generating functional for the 
two point function of a spinorial field of conformal weight 
$\delta=m+{d\over2}$. In fact, under 
$g_{ab}\,\to\, r^2 g_{ab}$, we have $q\,\to\, r^{-1}q$ and
$\chi_{q\eta}\,\to\, r^{-{d\over 2}}\chi_{q\eta}$, so that
$\Theta(x,y)\,\to\, r^{-2\left({d\over2}+m\right)}
\Theta(x,y)$.
Let us point out also that the free spinorial indices in the two point 
function are acted on by the local $Spin(d)$ group.

Again, in the case $M={\bf R}^d\cup\{\infty\}$, our result reduces to
the known ones \cite{henningson}.

\section{Conclusions}

As it has been shown above, there is a well defined way to 
generalize the AdS/CFT correspondence to the more general situation 
in which the AdS space is replaced by a filling in of a generic compact
manifold. This is due to the fact that the procedure of reduction to 
the boundary is arranged as a short scale phenomenon on the bulk
(see also \cite{susswit}).
At short distances from the boundary and locally, the manifold
$\tilde M$ can be approximated by a negative constant curvature
space and therefore the reduction to the boundary theory holds
as if one was on an AdS space. More concretely, one can consider
the geometry of $\tilde M$ near the boundary.
Let $\left(e_a,\omega_a^{\,\, b}\right)$ [resp. $\left(\tilde e_A,
\tilde\omega_A^{\,\, B}\right)$] be the vielbein and spin connection
on $M$ [resp. on $\tilde M$]: the following relations hold
(we used them also to calculate (\ref{diracop}))
$$
\tilde e_0={d\xi\over\xi}, \quad \tilde e_a= {1\over\xi} e_a,\quad 
\tilde\omega_a^{\, 0}=-{1\over\xi}e_a, \quad
\tilde\omega_a^{\,\, b}=\omega_a^{\,\, b} 
$$
and the structure equations on $\tilde M$ can be written as
$$
\tilde T_0=0\, ,\quad \tilde R_a^{\,\, 0}=-e_a\wedge
{ d\xi \over \xi^2}- {1\over\xi} T_a
$$ $$
\tilde T_a= {1\over\xi} T_a\, ,\quad
\tilde R_a^{\,\, b}= - {1\over\xi^2}e_a\wedge e^b+ R_a^{\,\, b} \, ,
$$
where $T_a$ and $R_a^{\,\, b}$ [resp. $\tilde T_A$ and $\tilde R_A^{\,\, B}$]
are the torsion and curvature 2-forms on $M$ [resp. on $\tilde M$].
In the limit $\xi\sim0$ the first term in the curvature 
$\tilde R_A^{\,\, B}$ dominates and
$$
\tilde R_A^{\,\, B}\,\sim\, -\tilde e_A\wedge\tilde e^B
$$
which means that locally near the boundary
\footnote{The precise meaning of ``near the boundary" is
$\xi<< |R|^{-1/2}$, where $R$ is the scalar curvature on $M$.}
the $\tilde M$ geometry is AdS (with some torsion if $T_a\not=0$).
This explains also the stability of the spectral correspondence 
with respect to changes in the geometry of the boundary manifold.

\vspace{2cm}

{\bf Aknowlodgements}: I would like to thank M. Bertolini, 
A. Hammou, F. Morales, C. Scrucca and M. Serone for discussions and 
friendness and in particular L. Bonora for a careful reading of the 
manuscript.

\end{document}